(Preprint)

# OPTICAL NAVIGATION FOR INTERPLANETARY CUBESATS

Stephen R. Schwartz,[*] Shota Ichikawa,[†] Pranay Gankidi,[‡] Nalik Kenia,[§] Graham Dektor[**] and Jekan Thangavelautham[††]

CubeSats and small satellites are emerging as low-cost tools for performing science and exploration in deep space. These new classes of satellite exploit the latest advancement in miniaturization of electronics, power systems, and communication technologies to promise reduced launch cost and development cadence. JPL's MarCO CubeSats, part of the Mars Insight mission, will head on an Earth escape trajectory to Mars in 2018 and serve as communication relays for the Mars Insight Lander during Entry, Descent and Landing. Incremental advancements to the MarCO CubeSats, particularly in propulsion and GNC, could enable these spacecraft to get to another planet or to Near Earth Objects. This can have substantial science return with the right science instrument. We have developed an interplanetary CubeSat concept that includes onboard green monopropellant propulsion system and that can get into a capture orbit around a neighboring planet or chase a small-body. One such candidate is the Martian moon Phobos. Because of the limits of current state-of-the-art CubeSat hardware and lack of an accurate ephemeris of Phobos, there will be a 2–5-km uncertainty in distance between the spacecraft and Phobos. This presents a major GNC challenge when the CubeSat first attempts to get into visual range of the moon. One solution to this challenge is to develop optical navigation technology that enables the CubeSat to take epicyclic orbits around the most probable location of the target, autonomously search and 'home-in' on the target body. In worst-case scenarios, the technology would narrow down the uncertainty of the small-body location and then use optical flow, a computer vision algorithm to track movement of objects in the field of view. A dimly lit small-body would be detected by the occlusion of one or more surrounding stars. Our studies present preliminary simulations that support the concept. The results point toward a promising pathway for further development and testing aboard a demonstrator CubeSat.

## INTRODUCTION

CubeSats and small spacecrafts are emerging as low-cost platforms for performing planetary science. This is thanks to miniaturization of electronics, sensors and actuators. This small platform may be targeted towards short, focused, science-led missions[1]. In the solar system, there are 150,000+ small bodies that range up to 10s of kilometers in size[2]. Constellations of CubeSats and small spacecraft could prove to be the most feasible means to explore these small bodies. CubeSats and small spacecraft can be carried as hosted-payload aboard a large mission or be deployed independently on an Earth-escape trajectory. NASA JPL's IRIS v2 radio offers the ability for the CubeSat to communicate and be tracked using the Deep Space Network (DSN)[3,4]. Our work in this field has identified CubeSats as an ideal platform for exploring small bodies, moons, and Near Earth Asteroids (NEAs)[5,6,7]. One such target is Phobos[5], others include asteroids such as ~34-m 2000 UK11. There are unique challenges to exploring these small bodies. Our knowledge of these small bodies is limited, the ephemerides are not accurate enough for mission navigation, and bodies are so small and dark that a spacecraft en route to the body may run the risk of flying past them without detection. Therefore, in many cases, optical navigation is critical in the search for, and the rendezvous with these targets.

---

[*] Postdoctoral Fellow, Space and Terrestrial Robotic Exploration Laboratory, Arizona State University, 781 E. Terrace Mall, Tempe, AZ..
[†] Master's Student, Space and Terrestrial Robotic Exploration Laboratory, Arizona State University, 781 E. Terrace Mall, Tempe, AZ.
[‡] Master's Student, Space and Terrestrial Robotic Exploration Laboratory, Arizona State University, 781 E. Terrace Mall, Tempe, AZ.
[§] Master's Student, School of Matter, Energy and Transport Engineering, Arizona State University, 781 E. Terrace Mall, Tempe, AZ.
[**] Master's Student, School of Matter, Energy and Transport Engineering, Arizona State University, 781 E. Terrace Mall, Tempe, AZ.
[††] Assistant Professor, Space and Terrestrial Robotic Exploration Laboratory, Arizona State University, 781 E. Terrace Mall, Tempe, AZ.

Optical navigation benefits from high-resolution visible and thermal imagers. Rapid advancements in these sensor technologies make them viable for use on CubeSats. A second technology useful for optical navigation is good, compact magnification systems that can fit a CubeSat. A third important technology is a propulsion system[14] with enough delta-v to perform a systematic search in the vicinity of the target asteroid. A fourth element is crucial: high-speed image processing using Field-Programmable Gate Arrays (FPGAs). FPGAs enable image processing using an array of logical gates. Due to their wide use in image filtering and pre-processing, FPGAs typically are built into scientific imaging platforms.

Combining all of these technologies, a spacecraft would get into the target-object's orbit around the sun or large body. This will be followed by a systematic visual/thermal scan of the surroundings. The scan will be performed to identify any change in the background star-field and occlusion or movement by a nearby body. The challenge is that the body, in the worst-case scenario, may be relatively dark/cold and appear only in one or very few pixels, making it nearly indistinguishable from the surrounding dark sky. Upon detection, the spacecraft would then use its visual sensors to perform a visual gradient descent search on the target. Utilizing its propulsion system, the spacecraft would home-in on and ultimately rendezvous with the target.

Advances in optical navigation using small spacecraft and CubeSats will open the door to exploring many small solar system bodies quickly and cheaply. Large missions require long periods of time to plan and execute, and are expensive. However, CubeSats and small spacecraft, owing to their small size and low-cost offer the best opportunity yet to access these small bodies. Our work identifies optical navigation as a critical enabling technology for the exploration of small bodies in the solar system. With the limitations of current ephemeris and limitations in DSN tracking, a small spacecraft or a CubeSat will need to use its onboard suite of sensors to navigate towards a very small target. We will first provide background to the problem and identify a candidate CubeSat for performing small-body exploration. Next, we will present our optical navigation algorithm and analyze the image processing algorithms to detect small bodies before providing a discussion. This is followed by a section on conclusions and future work.

**SYSTEM OVERVIEW**

Inspired by the capabilities of NASA JPL's Mars Cube One (MarCO)[8,9], we have been designing interplanetary CubeSats that can be deployed as hosted payloads, mother-daughter crafts and free-flyers. Figure 1 shows a free-flyer CubeSat design called LOGIC, initially developed for a Phobos mission[5]. The spacecraft is also well suited for flyby and rendezvous missions to small bodies. Table 1 shows the mass and volume budget. The spacecraft contains two science instruments, namely the e2V Cires visible camera and the FLIR Tau 2 thermal camera. The spacecraft is powered using a pair of onboard deployable solar panels called eHawk+ from MMA Design.

The solar panels have 1 degree of freedom gimballing. The backside of each solar panels contains an X-band reflectarray for communication. The spacecraft will charge two Gomspace NanoPower BPX lithium ion batteries with a total energy capacity of 150 Whr.

During the insertion burn and periodic communication transmission to Earth, the spacecraft will require use of both the solar panels and onboard batteries to generate the required power. If the solar-panel deployer fails, the mission will not be salvaged. This is due to the high power demands by the communication system.

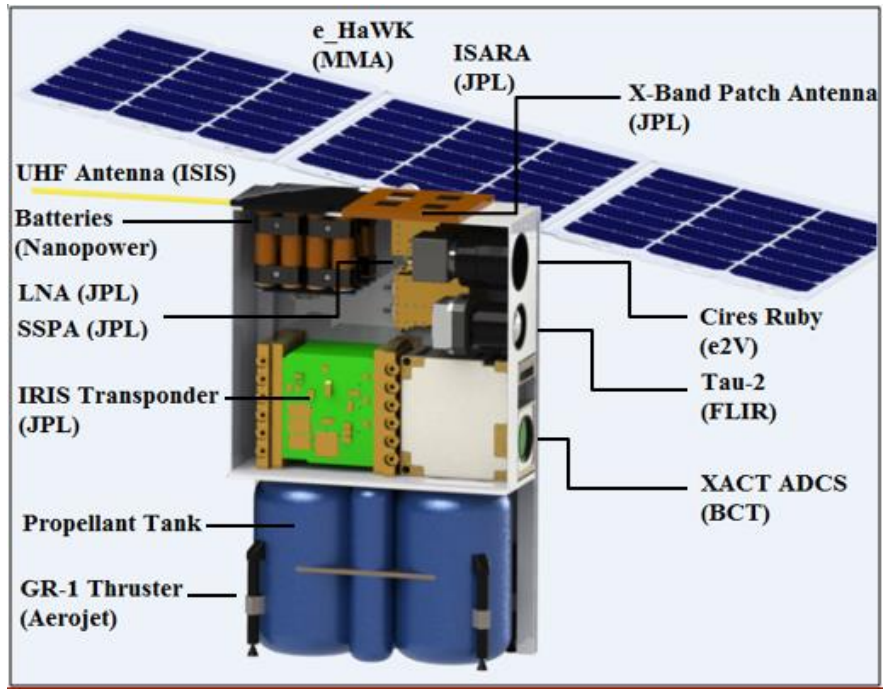

**Figure 1: Layout of a CubeSat for interplanetary and small-body exploration.**

**Table 1: Mass and Volume budget**

| System | Mass (kg) | Internal Volume (cm$^3$) |
|---|---|---|
| Chassis | 1.0 | - |
| Communications | 2.2 | 1,100 |
| Power | 1.9 | 700 |
| Thermal | 0.1 | 50 |
| ADCS | 0.9 | 500 |
| Propulsion | 5.1 | 3,400 |
| Payload | 0.45 | 350 |
| **Total** | **11.7** | **6,100** |
| Requirement | 14.0 | 8,000 |
| **Margin** | **17%** | **24%** |

The spacecraft contains a green monopropellant system developed by Aerojet Rocketdyne that can be used for flyby, rendezvous, or insertion. This propulsion system consists of four 1-N GR-1 thrusters. This system allows the spacecraft to get into a capture orbit, perform trajectory corrections, and desaturate the reaction-wheels. The system contains 3.3 kg total fuel and occupies 1.8 kg of component mass. The spacecraft propulsion system has an estimated[5] delta-v of up to 830 m/s. The remaining fuel allows for desaturation of the reaction wheels, correction maneuvers, and orbit optimization.

The spacecraft utilizes the rad-hardened SpaceCube MINI from Aeroflex as its main command and data-handling computer. The SpaceCube MINI consists of a primary Xilinx Virtex-5QV space-qualified FPGA processor and a daughter board with Aeroflex UT6325 FPGA. Use of FPGA processors enables high-level operational flexibility. In addition, FPGAs, due to their distributed architecture, are better suited for high-radiation environments and for optical navigation. The spacecraft uses the DSN for communication, as well as and Doppler tracking using the JPL IRIS v2 radio (MarCO heritage)[3]. The radio is capable of up to a 250 Kbps data rate depending on available power. In addition, the spacecraft contains a UHF radio for backup communication with nearby assets.

**Primary Instruments**

To meet the typical science and exploration goals, the CubeSat will utilize a Commercial Off-the-Shelf (COTs) assembled thermal camera (Figure 2) and visible camera (Figure 3)[5]. The instruments, as will be shown, play the dual purpose as a science instrument and navigation instrument for small-body exploration. Trade studies were performed to finalize on these two science payloads[5]. The thermal camera is a FLIR Tau 2 with an NFOV lens and a Cameralink module.

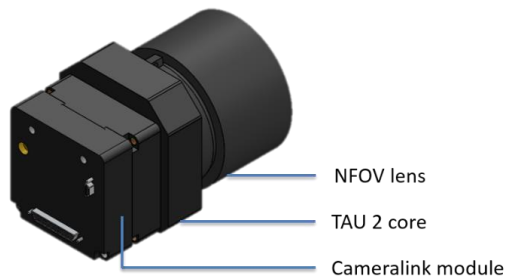

**Figure 2: Thermal camera assembly**

The visible camera is composed of E2v's Cires camera with a lens system based on the Pentax B5014A lens (Figure 3).

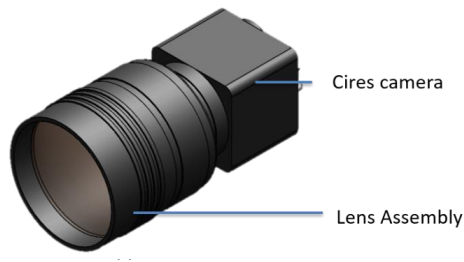

**Figure 3: Visible Camera assembly**

The product specifications and performance of both instruments have been summarized in Tables 2 and 3.

Table 2: Thermal Camera Performance Specifications[10,11].

| Characteristic | Value |
| --- | --- |
| Assembled mass (g) | 200 |
| Assembled volume (cc) | $7.5 \times 4.5 \times 5$ |
| Power (W) | 1.2 |
| Voltage (V) | 5 |
| Operating temp. (K) | 233 – 353 |
| NEΔT at ambient (mK) | 50 |
| AFOV | 18 x 14 |
| Pixel Density | 640 x 512 |
| Integration Time | 5 |
| Spatial res.@50 km (m/px) | 24.75 |

Table 3: Visible Camera Performance Specifications[12,13].

| Characteristic | Value |
| --- | --- |
| Assembled mass (g) | 250 |
| Assembled volume (cc) | $9.4 \times 4.8 \times 4.8$ |
| Power (W) | 1.5 |
| Voltage (V) | 12 |
| Operating temp. (K) | 253 – 323 |
| SNR (dB) | 39 |
| DNR (dB) | 65 |
| QE (%) | 60 |
| AFOV | 7.76 x 6.18 |
| Pixel Density | 1280 x 1024 |
| Frame rate (FPS) | 60 |
| Spatial res. @50 km (m/pixel) | 5.29 |

**OPTICAL NAVIGATION**

An onboard visual camera would assist in locating the asteroid or small body, but the search would be planned around the thermal camera, which has the following resolution at the following distances (Table 4). The angular resolution is 1.7 arcmin/pixel. For our purposes here, we consider a nominal asteroid with the size and semimajor axis of 2000 UK11, which is an estimated 34-m across. To explicate our navigation concept, we assume a simple non-inclined, zero-eccentricity orbit at a = 0.883 au. The navigation procedure is easily adapted to other orbits. If we consider the target to have an orbital uncertainty of a few arc-seconds due to runoff, this would translate to an uncertainty in position of a few thousand kilometers. Our task is to sweep the area, locate our target, and rendezvous.

**Table 4: Object Resolution Using Tau 2 Thermal Camera**

| Distance of craft from object (km) | Resolution (m/pixel) |
|---|---|
| 70 | 34.65 |
| 50 | 24.75 |
| 30 | 14.85 |
| 10 | 4.95 |

The onboard processor compares the successive images of the sky that it generates, performs the standard data reduction/cleaning routines, searching for objects by subtracting out the background star-field, effectively scanning for motion with respect to the background star-field or for points with periodic changes in flux, which would potentially indicate the relative motion or the rotation, respectively, of a small body.

The craft's ability to detect thermal variations between frames is limited by the thermal camera's IR-flux detection limit, while its ability to discern size is limited by pixel resolution. We estimate that a 5-second exposure time is sufficient for these purposes[5].

Figure 4 shows the co-moving orbital frame, centered on the expected location of the asteroid (initially the most probable location), $r_0$. The spacecraft is trailing this location by $s$, and matching its orbital velocity, $v_0$. In this case, the spacecraft makes an excursion from its initial position (in this reference frame of the co-moving orbit) from trailing $r_0$ by $s$, to passing it at an orbit $s$ interior to $r_0$, to leading it by $s$, to being passed by it while on an orbit $s$ exterior to $r_0$, and then back to its original position with respect to $r_0$, once again trailing it by $s$ (this gives the orbit a diamond shape in the co-moving frame). The trip in Figure 4 takes 10 hours, with $s$ = 30 km, 4 thrusts of 15 seconds each, and, assuming 14 kg, required us to spend the fuel shown in the Figure 5 (~100 grams). We can perform this maneuver more slowly or more quickly depending on mission requirements (more uncertainty entails slower sweeps to conserve fuel), with the required fuel consumption varying roughly inversely with time and linear with $s$, so long as $s <<$ solar orbit. Effectively, we have the spacecraft taking epicyclic orbits around the most probable location of the target, and we take this diamond-shape trajectory for its simplicity, but other trajectories may be taken to minimize fuel consumption or for other case-specific considerations.

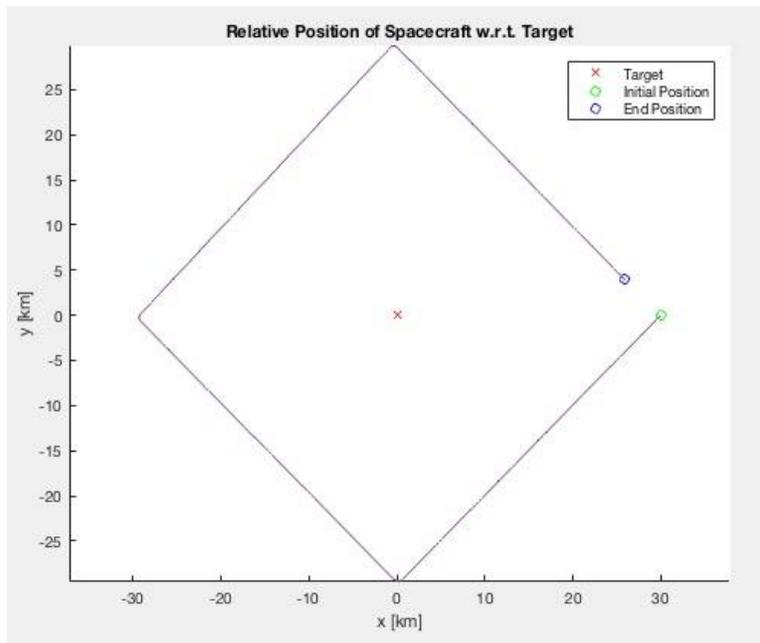

**Figure 4: Trajectory of 'diamond maneuver' performed to search for target small-body.**

All the while, we scan the sky. The FLIR Tau 2 thermal imager provides 14.85 m/pixel @ 30 km (1.702 arcmin). The detector array is 640 × 512 pixels, thus each image looks at 0.08029 sr, or 1/156.5 of the total sky. Each IR image is exposed over 5 seconds. We conservatively grant an additional 1 s between exposures to smoothly reposition about 15º to the next frame and to allow vibrations caused by the motor to damp out. Thus we image the entire sky approximately every 15m 40s (939 sec).

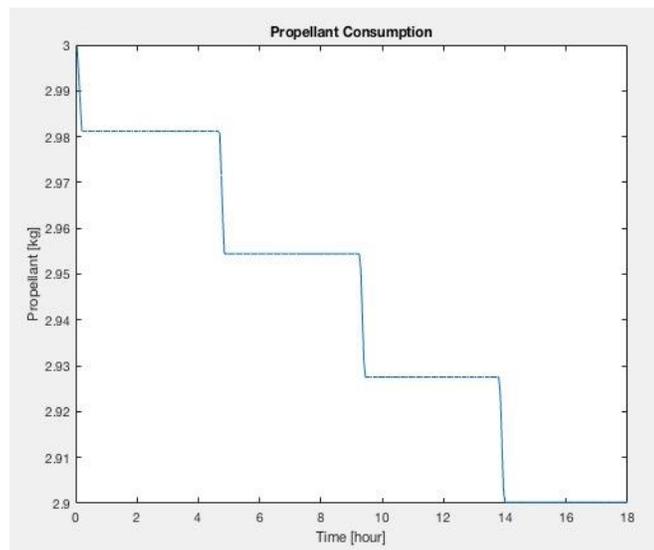

**Figure 5: Propellant consumed for performing a 'diamond' maneuver.**

A proposed scanning strategy begins with successfully navigating to the target's orbit, as close to its anticipated position as possible, $r_0$, and achieving velocity $v_0$. (1) stay in this location and perform several full-sky scans, taking 15m 40s each. (2) perform a maneuver to put the craft in a position that trails $r_0$ by $2s$, with a velocity matching the orbital velocity (requires simple thrust against the direction of motion, then after moving $2s$, an equal and opposite thrust (in the direction of motion) to park $2s$ behind $r_0$ (decay is minimal as $s \ll$ orbital radius). (3) perform a diamond-shape maneuver as shown in Figure 4, but with a baseline of $4s$ instead of $2s$ (scanning all the while). (4) perform the same maneuver, but out of the orbital plane (orthogonal). (5) move to a location $r_0 - 4s$, perform a "diamond" maneuver with an $8s$ baseline, do another maneuver at 45° out of the plane, then another orthogonal (90°) to the orbital plane, than another at 135°, then move to location $r = r_0 - 6s$. The loop breaks when the spacecraft detects a potential target body and goes into a confirmation routine (6); it thrusts toward the target body upon confirmation (7).

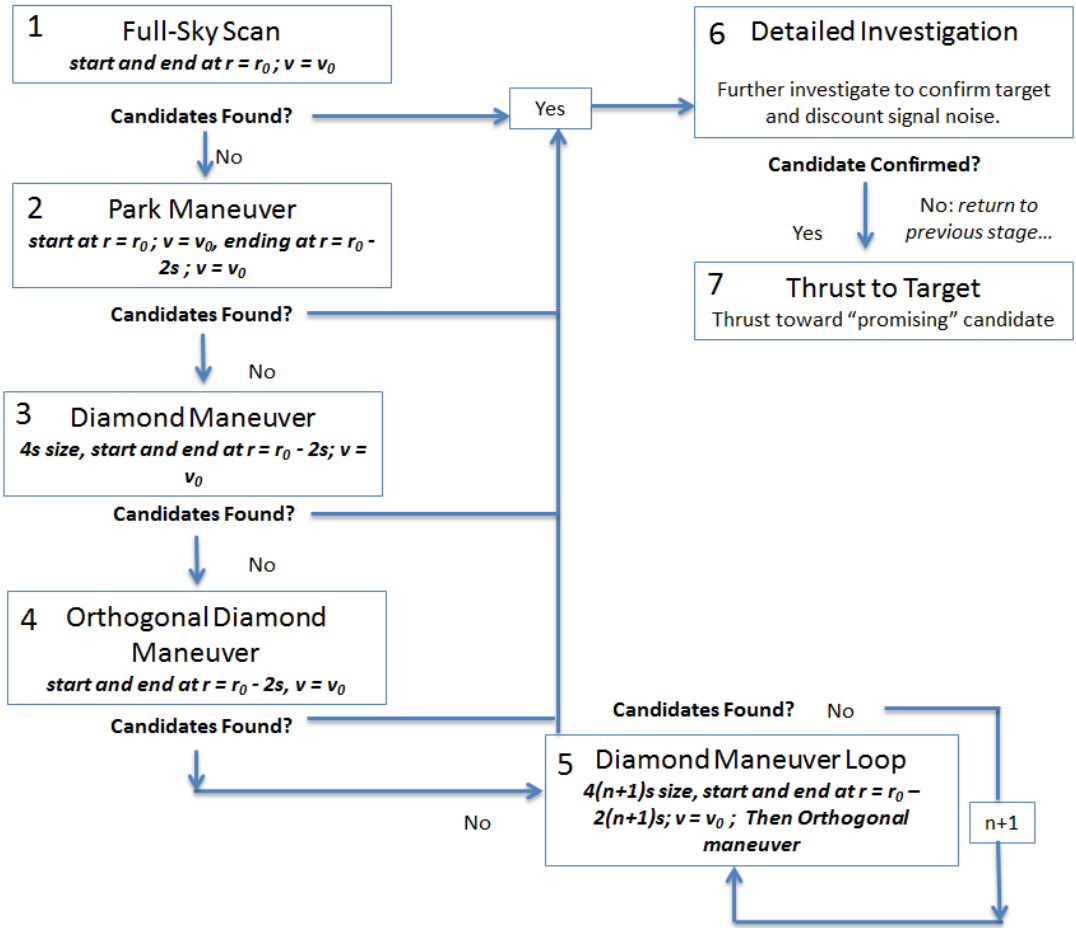

**Figure 6.** Visual navigation search algorithm to get to target small-body.

### Covering all volume

We require the detection range to be somewhat greater than $s$ so that we can cover regions sufficiently above or below the orbital plane. To be conservative, we require this detection range to be at least $\sqrt{2}s$. Additional out-of-plane maneuvers are required as we go to larger distances to account for locations above or below the orbital plane, assuming similar uncertainty in all three dimensions.

**Detecting rotation-induced thermal flux variation**

The period of the full-sky scans, $P_{skyscan}$, should be checked against the expected rotation period of the asteroid, $P_{rot}$. If these are close to integer multiples of each other (unless $P_{rot} \gg P_{skyscan}$), this could prevent detection via periodic changes in flux from the object caused by its rotation. If the rotation rate is unknown or highly uncertain, it may make sense to consider varying $P_{skyscan}$ in some semi-irregular fashion to ensure that we allow detection by periodicity in flux (asteroid rotation).

A *confirmation routine* pauses the all-skyscan to stare at each point in question for some period of time, seeking detection. If nothing is detected, the craft would revert back to its prior place in the broad search routine, or, if there is reason to believe that the object may have strayed, the craft would start to scan around this region of the sky. It would do so by spiraling out its view around the expected location in the sky to some specified extent before going back to the broad search (if not found).

If one or more objects were to be confirmed, the craft would then navigate toward the suspected region of sky that matches best what is known about the object being sought. The craft matches the angular speed, with respect to the camera, of the thermal anomaly (such that background stars move in image frame, object stays fixed in frame). The spacecraft thrusts toward it, keeping the object fixed in center of image frame. An increase in the flux and/or size of the thermal signature of the candidate would be expected if the object were near. Meanwhile, estimates of the object's size, distance, spin, albedo, etc. are refined. When the craft arrives at the object, it then enters the next phase, its post-rendezvous operations (surveying, imaging).

**Detection of the small-body**

Detection of an asteroid or small body using a visible and thermal camera is a challenge. In a nominal scenario, the asteroid or small-body appears as a small, dark object that spans perhaps only a few pixels. While scanning, the asteroid would appear much like another faint star, but with a thermal signature characteristic of an asteroid. Upon staring at the object (Step 6 of Figure 6), this object will move in a different direction with respect to the stationary background stars identified in our star catalog.

The asteroid or small body in the worst-case scenario occupies 5 pixels or less, is dark and emanates little or no thermal signature. As there are no features to discern, it is impossible to use feature detection algorithms to match or scan for the small body. Instead, our approach uses the background star field to build the star catalog as with the nominal scenario.

Figure 7 shows a background star-field. As the spacecraft is panning, we would use the standard optical flow algorithm to identify and catalog the brightest stars. Optical flow is used to identify motion of objects in a visual field. Consecutive images are compared to find motion vectors. As the image is panning, the stars should all show relative motion vectors that are identical in direction and magnitude (see Figure 8). A threshold is applied to the motion vectors, so vectors below the threshold are ignored due to effects of camera noise and above the threshold are kept. This threshold filtering reduces the number of motion vectors that need to be analyzed. Furthermore, this approach simultaneously enables identifying the brightest stars (see rectangle in Figure 8) while the camera is in motion, but without blur. When a bright star is occluded by the asteroid (see Figure 9), the resultant optical flow vector in Figure 10, shows the star missing (black rectangle). By a process of elimination, we have identified the location of the asteroid in the image. Upon detection, a visual confirmation routine will be performed. The spacecraft would then use its image to perform a visual gradient descent search on the target and use its propulsion system to home-in on the target small body.

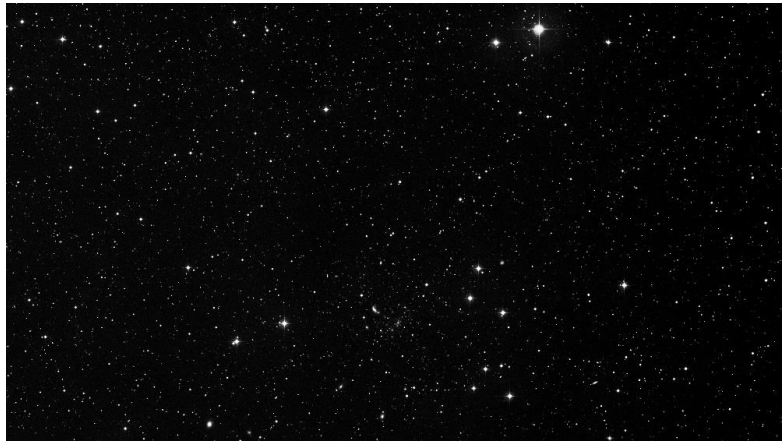

**Figure 7. Visible image of surrounding star field.**

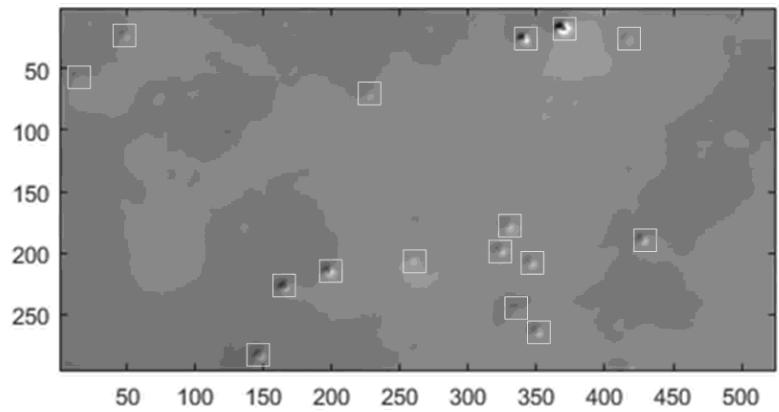

**Figure 8. Resultant stars cataloged from star field shown in Figure 7 using optical flow.**

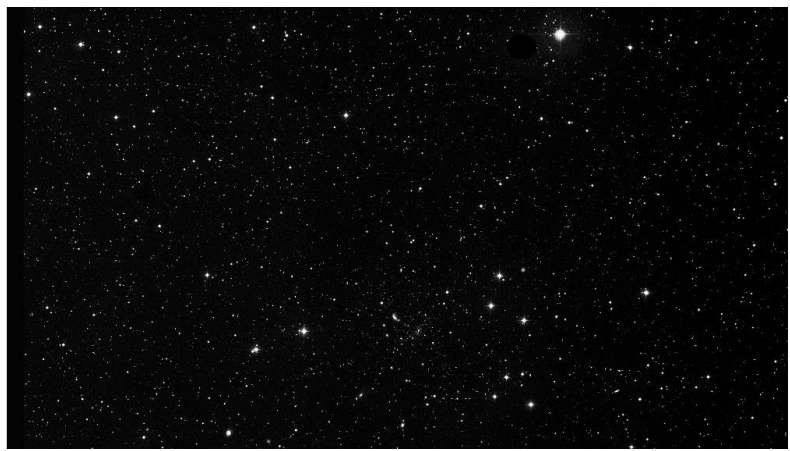

**Figure 9. Visible image of surrounding star field, panned 15 degrees, with occlusion of a star by the target small-body.**

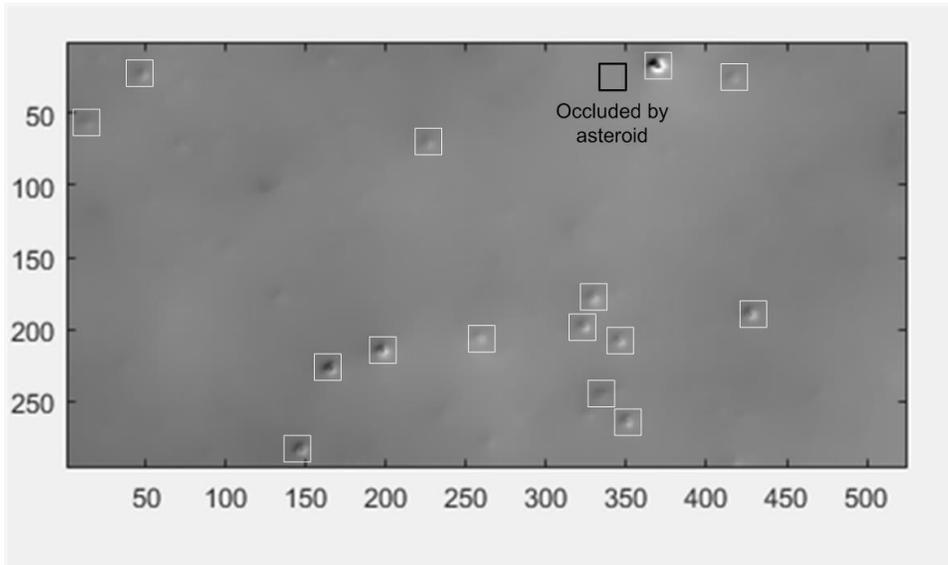

**Figure 10. Resultant stars cataloged from star field shown in Figure 9 using optical flow. A star occluded by the asteroid (target small body) has been identified.**

For us to locate the asteroid quickly requires a sufficient catalog of stars closely spaced and distributed throughout the field of view. This enables us to quickly identify an instance of occlusion and thus pinpoint the asteroid. The example optical flows images from Figures 8 and 10 require a lower threshold to be used for practical purposes, as the cataloged stars are not well distributed throughout the field of view. Using this optical-flow approach, we identify the asteroid by a process of elimination when comparing relative motion of the background star-field. It enables us to use a slow moving camera to make the determination, thus achieving minimum scan-time of the surrounding. Any faster movement will result in motion blur. Optical flow is a simple machine vision technique that can readily be implemented on FPGAs accompanying current thermal and visible imagers. This should provide real-time detection, due to the parallel processing that is possible with the hardware. Thus the processing and identification of the asteroid can all be done using the imaging hardware and without taxing the spacecraft main computer.

**CONCLUSION**

Exploration of small bodies, 10s of km down to few meters using small spacecraft and CubeSats presents a new opportunity in exploration. There are tens of thousands of these asteroids in the vicinity of Earth and many more in the main asteroid belt. These asteroids typically have low-albedo and are hard to detect with even the best imaging systems. Current ephemerides of these small bodies are inaccurate or incomplete. This makes it especially challenging to navigate to these target small bodies. In our approach, we propose use of interplanetary CubeSats that would be dropped on an Earth escape trajectory and perform an insertion burn into a nearby orbit to the target body. The CubeSat would use its onboard visual and thermal imagers to systematically scan the vicinity and by pursuing a 'diamond' shaped trajectory around the target body. If the target body is not found, the spacecraft would enlarge the search space and fully scan the surroundings until it can home-in on the target asteroid to perform detailed science observations. The proposed approach uses optical flow to identify the target small-body that occludes back-ground stars. The asteroid detection algorithm can be implemented entirely in hardware using current FPGAs accompanying thermal and visible imagers. Work is proceeding on implementing the algorithm and verification on laboratory hardware.

**ACKNOWLEDGEMENT**

The authors would like to thank Prof. Erik Asphaug for his constant encouragement in pursuing a mission to a small-body. In addition, the authors would like to thank NASA JPL's Dr. Andrew Klesh, Dr. Alessandra Babuscia, Dr. Julie Castillo-Rogez and JPL's Team Xc for advancing the LOGIC CubeSat concept. The authors would like to gratefully acknowledge to JPL's SURP Program and JPL's Office of the Chief Scientist for graciously organizing a visit to JPL